\documentstyle[preprint,aps,epsf,floats]{revtex}
\def\ada{\left({\dot{a}\over{a}} \right)^2}
\def\adda{{\ddot{a}\over{a}}}
\def\A0{A_0(y)}
\def\warp{e^{-2A_0(y)}}
\begin{document}

\preprint{\tighten \vbox{
                \hbox{CMU-HEP00-03}
                \hbox{} }}

\title{Cosmological Solutions on Compactified $AdS_5$ \\ with a Thermal Bulk}

\author{I.\ Z.\ Rothstein}

\address{\vspace{.5cm}
 Department of Physics, 
Carnegie Mellon University, Pittsburgh, PA 15213 }

\maketitle

{\tighten
\begin{abstract}

This paper is an investigation of the effects of a thermal bulk 
fluid in
brane world  models compactified on $AdS_5$. 
Our primary purpose is to study how
such a fluid changes the bulk dynamics and to compare these
effects
with those generated by matter localized to the branes. 
We find an exact cosmological
solution for a thermally excited massless bulk field, as well
as  perturbative solutions with matter on 
the brane and  in the bulk. 
We then perturb around these solutions to find solutions
for a massive bulk mode in the limit where the bulk mass ($m_B$) is small
compared to the AdS curvature scale and $T< m_B$.  We find that without a stabilizing
potential there are no physical solutions for a thermal bulk fluid. We then
 include a stabilizing potential and calculate the shift in the
radion as well as the time dependence of the weak scale as a function
of the bulk mass. It is shown that, as opposed to a brane fluid, 
the bulk fluid contribution to the bulk dynamics is controlled
by the bulk mass. 
\end{abstract}
}

\newpage

\section{Introduction and Motivation}
The possibility that we are living on an isolated defect has
led to new paradigms which are now being explored. While the low
energy effective theory of these models 
correctly reproduces the standard model physics, we expect that the
cosmology should be quite dissimilar at large enough temperatures. In particular, the cosmology of
models which solve the hierarchy problem through warped compactification, such as 
the Randall Sundrum model\cite{RSI} (RSI), are especially
interesting because we expect to see deviations from the standard
cosmology  as we approach the weak scale\cite{Binetruy,Kaloper,Nihei,CsakiI,ClineI,CsakiII,ClineII,Mohapatra,Kyae,Kim,Lesgourgues,Grinstein,KantiI}.
The cosmology of these models has been discussed previously including
the effects of matter density on the branes. 
What makes the cosmology of these models particularly interesting is
that we expect five dimensional gravity to become strongly coupled  at
 the TeV scale. Above the TeV scale, it is believed that the theory
should still have a description in terms of  a local quantum
field theory, via the holographic principle \cite{holo}. 
In this paper we will continue the exploration of the cosmology
of (RSI) by studying the effects of a thermal fluid in the bulk.
We will study
temperatures low enough so that, $T<m_{kk} M_5/k$ (where $M_5$ is the five
dimensional Planck scale and $k$ is the $AdS$ curvature scale)
and a weakly coupled ($k<M_5$) five dimensional gravity description is still valid. 
Since our results only depend upon the geometry they are also applicable
to models where the hierarchy is stabilized via supersymmetry  and generated via the geometry \cite{susy}. However, we will couch our results 
in terms of (RSI).

The primary motivation for this work stems from the 
fact that there must
be some nontrivial bulk physics in the RSI model, as well as in 
the supersymmetric models mentioned above.
Thus, we would expect that at some point in
its thermal history, more specifically at the TeV scale,  a thermal bulk will be relevant.
 In its purest form, RSI entails the trapping
of the standard model particles on a brane of negative tension
positioned at an orbifold fixed point. At some finite proper distance
in the extra dimension there is a positive tension brane where
 the graviton is localized. In this choice of coordinates, 
the small overlap of the graviton 
wave function with the standard model brane leads to the apparent
weakness of gravity. This minimal model must be augmented to account
for the stabilization of configuration. That is, in the minimal RSI,
the distance between the branes parameterizes a flat direction in
field space. The introduction of new physics in the bulk can lift this flat
direction. For instance, the 
Goldberger-Wise (GW) mechanism is a natural way to lift the degeneracy
by introducing a bulk scalar with a non-trivial vacuum profile. 
Thus, in general, we may expect that in addition to gravitons propagating in
the bulk, there will be other fields which may play a role in
the bulk physics. In fact, it is phenomenologically viable to
have particles with standard model gauge charges propagate in
the extra dimension\cite{models}. If this were the case then bulk thermal effects
would be enhanced due to the large number of species which would
equilibrate.

In this work we seek to understand how the cosmology changes
once the bulk thermal effects are taken into account.
We would expect that at long distances the bulk fluid will have
the same effect as a fluid trapped to the brane. Though we might be
concerned by the fact that, since the lapse function is changing
exponentially, the effect of the bulk fluid at long distances
could differ from those of a brane fluid.
It has been pointed out that    the visible
and Planck brane matter densities ($\rho,\rho_*$)
contribute to the expansion rate as \cite{CsakiII}
\begin{equation}
H^2=\frac{8\pi G_N}{3}(\rho_\star+\rho e^{-4kb_0}),
\end{equation}
where $b_0$ is the proper distance between the branes. Given that
the natural scale of the Planck brane is $M_{pl}$, this puts strong
constraints on the brane matter density, as pointed out in
\cite{CsakiII}. Indeed, this constraint implies that the Planck brane
must be in its vacuum state for all intents and purposes. 
Thus, depending on the bulk matter density profile, a
bulk energy density  could
lead to  over-closing the universe.

We would also expect a bulk fluid to effect the bulk dynamics quite
differently from a brane fluid. A brane fluid generates a non-trivial
radion potential, which in the absence of a stabilizing mechanism, 
tends to push the branes apart, unless one imposes the unphysical
fine-tuning between the visible brane energy density ($\rho$) and
Planck brane density $(\rho_\star)$, $\rho_\star=-\rho e^{-4kb_0}$\cite{Binetruy,Kaloper,Nihei}.
  Here we calculate the radion potential  generated
by a bulk thermal fluid \cite{Brevik}, and determine its effect on  the above fine tuning relation. We then include a stabilizing potential to see how
a bulk fluid shifts the radion 
expectation value and the weak scale as  functions of time.

To address  these issues we will  solve the coupled
bulk fluid Einstein equations. We start by finding an exact solution
with fluid in the bulk and vacuum branes. 
In general, this is rather intractable  problem, since the sources
on the right hand side of the Einstein equation are complicated
highly non-linear functions of the metric ansatz. However,  we will be able to
find an exact  solution in  the particularly simple case of  a massless field
at temperatures too low to excite higher Kaluza-Klein modes. We then find a realistic
solution with matter in the bulk, as well as on the branes, which is
valid when the energy densities are small compared to the bulk
curvature (i.e. visible brane temperatures less than a TeV). 
We then perturb around the massless solution to find more
interesting solutions
which include the effects of a massive bulk field. 
We show that without a stabilizing mechanism, there are no physical
solutions. Furthermore, once stabilization is taken into account,
all the non-trivial bulk dynamics, e.g. the evolution of the weak
scale, are driven by the bulk field mass.

\section {Exact Solution with a Bulk Fluid}

The five dimensional action is given by
\begin{equation}
\label{5Daction}
L= \int d^5x \sqrt{-G}(-M_5^3R-\Lambda+L_B)+
\int d^4x \sqrt{-g}~L_{TeV}+\sqrt{-g_\star}~L_{Pl},
\end{equation}
where $g$ and $g_\star$ are the induced metrics on the visible
and Planck branes, respectively. $M_5$ is the five dimensional Planck
scale and  
$L_B$ is the bulk field Lagrangian which describes the stress
energy of the bulk fluid.
The fifth dimension is 
compactified on $S_1/Z_2$, with the
Planck and visible branes placed at the orbifold fixed points,
$y=0,1$, respectively.
For this section we will leave the branes empty, and thus, 
their only contribution to the action are due to their tensions
$V_\star$ 
and $V$.

We will search for solutions which have a stable  (time independent)
fifth dimension.
We thus make the following ansatz for the metric.
\begin{equation}
\label{exactansatz}
ds^2=e^{-2A_0(y)}dt^2-a(t)^2 e^{-2A_0(y)}d \vec{x}^2-b_0^2 dy^2.
\end{equation}
We could rescale the $y$ coordinate to eliminate the constant
$b_0$, but choose not to for bookkeeping purposes.
The  bulk stress energy tensor is
\begin{equation}
T_{mn}=\Lambda g_{mn}+g_{mp}T^{p}_{n},
\end{equation}
with $T^p_n={\rm diag}(\rho_B(t,y),-\vec{P}_B(t,y),-P_5(t,y))$.

The Bianchi identity leads to 
\begin{equation}
\label{bianchiy}
A_0^\prime(y)(\rho_B(t,y)-3P_B(t,y)+4P_5(t,y))-P_5^{\prime}(t,y)=0.
\end{equation}
\begin{equation}
\label{bianchit}
\dot{\rho}_B(t,y)+3 \frac{\dot{a}}{a}(t)\left(P_B(t,y)+\rho_B(t,y)\right)=0.
\end{equation}
Primes denote derivative with respect to $y$, and dots 
denote derivative with respect to the time-like coordinate.

The $G_{00},~G_{ii}$ and $G_{55}$ Einstein equations in the bulk are 
\begin{equation}
\label{G00B}
\ada-\frac{1}{b_0^2}e^{-2\A0}\left(2\left(A_0^\prime(y)\right)^2-A_0^
{\prime \prime}(y)\right)=\frac{\kappa^2}{3}exp^{-2A0(y)}(\rho_B(t,y)+\Lambda)~, 
\end{equation}
\begin{equation}
\label{GiiB}
\frac{3}{b_0^2}e^{-2A_0(y)}
(2(A_0^\prime(y))^2-A_0^{\prime \prime}(y))
-\left( \ada\!\!\!\!(t)+2\adda(t) \right)=\kappa^2 e^{-2A_0(y)}(P_B(t,y)-\Lambda)~,
\end{equation}
\begin{equation}
\label{G55B}
6(A_0(y)^\prime)^2-3b_0^2 e^{2A_0(y)}\left(\ada\!\!\!\!(t)+\adda(t)\right)=
\kappa^2b_0^2(P_5(t,y)-\Lambda)~,
\end{equation}
and we have defined $\kappa^2=\frac{1}{2M_5^3}$.
Combining the $G_{00}$ and $G_{ii}$ equations, we see that the bulk energy
density
and pressure must be of the highly restricted  form 
\begin{equation}
\label{form}
\rho_B(t,y)=\hat{\rho}(t)e^{2A_0(y)},~~P_5(t,y)=\hat{P_5}(t)e^{2A_0(y)}~~P_B(t,y)=\hat{P}_B(t)e^{2A_0(y)}.
\end{equation}
This restriction results from our simple choice of ansatz.
Whether or not a bulk fluid will indeed yield this form 
will be discussed later.

To solve this system of equations we first make 
 linear ansatz for $A_0$,  since the geometry should reduce
to AdS in the limit of vanishing bulk energy-momentum, which  leads to the usual result \footnote{From here on, the absolute value, implied by the orbifolding,  will be dropped
for notational simplicity.}$A_0=kb_0\!\!\!\mid \!\!\! y \!\!\!\mid  $
with $k=-\kappa^2\Lambda/6$, as in the RSI solution.
This result has been derived  using the $G_{00}$ and $G_{ii}$
equations. We have not, to this point,  imposed the $G_{55}$ equation which
for this simple ansatz is identical to the Bianchi identity
\begin{equation}
2\hat{P}_5(t)+\hat{\rho}_B(t)-3\hat{P}_B(t)=0,
\end{equation}
With the result that 
\begin{equation}
\label{bulkrho}
\hat{\rho}_B(t)=\frac{3}{\kappa^2}\ada\!\!\!\!(t),
\end{equation}
\begin{equation}
\hat{P}_B(t)=-\frac{1}{\kappa^2}\left(\ada\!\!\!\!(t)+2\adda(t) \right),
\end{equation}
\begin{equation}
\label{bulkP5}
\hat{P_5}(t)=-\frac{3}{\kappa^2}\left(\ada\!\!\!\!(t)+\adda(t) \right),
\end{equation}
which is seen to obey (\ref{bianchit}).
A related  solution was found previously in  \cite{Kyae,Kim}.
Note that the fluid tends to build up near the visible brane.
It is straightforward to see that the usual cosmology on the brane results
from this solution.

We must now deal with the fact that our ansatz leads to the constraint (\ref{form})
on the form of the bulk pressures and energy density. One simple 
case for which we will have a solution is vacuum domination, 
as discussed in \cite{Kyae,Kim}, where the constraints (\ref{form}) 
are obviously satisfied. Here we are more interested in the more
inevitable case
where the energy density arises from thermal fluctuations. 
In this case we must calculate the local pressure and energy density
using our solution to determine if the constraint is indeed satisfied.
We will postpone this calculation until  we have included the
effects of matter on the branes.

Before closing this section we note that  
the jump conditions 
\begin{equation}
3[A_0^\prime(0)]_{-} = \kappa^2V_\star~~~~-3[A_0^\prime(1)]_{-} = -\kappa^2V,
\end{equation}
lead to the usual fine tunings. One tuning is needed to
fix the flatness of the brane and the other to enforce the zero force
condition between the branes,
\begin{equation}
V_\star=6k/\kappa^2,~V=-V_\star.
\end{equation}
The fact that there are two fine-tunings is a consequence of the
lack of a radion potential for this solution, 
which one might have expected to be
generated by the existence of the bulk fluid. 
This result  is clearly related to the constraint (\ref{form}).
We shall explore this point further below.

\section{Solutions with Matter on the Brane and in the Bulk}

Solutions with matter on the brane were previously found in 
\cite{CsakiII,Lesgourgues,ClineII}. Here we  augment these
results to include the effects of a massless bulk fluid. The derivation is
similar to that in \cite{CsakiII} but is included for completeness.

The previous ansatz is not general enough to allow for
a solution with brane matter. We thus introduce the brane stress energy tensors
\begin{equation}
T^A_B=\frac{\delta(y)}{b_0}{\rm diag}(V_\star+\rho_\star,V_\star-p_\star,V_\star-p_\star,V_\star-p_\star,0)+\frac{\delta(1-y)}{b_0}{\rm diag}(-V+\rho,-V-p,V-p,V-p,0),
\end{equation} 
and we modify the ansatz, as follows \footnote{In \cite{Lesgourgues} a 
solution was found which is accurate to all orders in the  brane matter density.
However, the solution is only valid as long as
the energy density on the brane is small enough that it can still be
assumed that the $G_{55}$ Einstein equation is automatically satisfied. This
will be true only as long as the matter energy density is small
compared to the scales in the stabilizing field potential.}
\begin{equation}
\label{pertansatz}
ds^2=\exp^{-2A_0(y)}\left(1+2 f(t,y)\right)dt^2-
a(t)^2\exp^{-2A_0(y)}\left(1+2 g(t,y)\right) d \vec{x}^2-b_0^2 dy^2,
\end{equation}
and treat both the bulk and brane matter as perturbing sources around
the standard RSI solution.
 We will calculate to leading non-trivial
order in the perturbations $\rho\sim O(\epsilon)$.
Since $\dot{\rho}\sim \frac{\dot{a}}{a} \rho$, it will often be the case
that we will be able to drop time derivatives of the fluctuations.
 The solutions will
again restrict the bulk profile of the matter density to be of the
form (\ref{form}).

The $G_{05}$ Einstein equation is now no longer automatically satisfied,
and
is given by
\begin{equation}
\label{G05}
\dot{f}^\prime(t,y)+\frac{\dot{a}}{a}(t)\left(f^\prime(t,y)-g^\prime(t,y)\right)=0.
\end{equation}
In this equation we must keep the time derivative of the perturbation, since
it is leading order.
While the $(00),~(ii)$ and $(55)$   linearized  Einstein equations are
given by

\begin{eqnarray}
\label{G00f}
3\ada e^{2A_0(y)} -6(A^\prime_0(y))^2+12A^\prime_0(y)f^\prime(t,y)- 
3f^{\prime \prime}(t,y)=
 b_0^2\kappa^2(\warp \rho_B(t,y)),
\end{eqnarray}
\begin{eqnarray}
\label{Giif}
-4\frac{k}{b_0}(2f^\prime(t,y)+g^\prime(t,y))+
\frac{1}{b_0^2}(2f^{\prime
\prime}(t,y)+g^{\prime \prime}(t,y))-e^{2A_0(y)}\left( 2\adda(t)+\ada\!\!\!\!(t)
\right)=  
\kappa^2 (P_B(t,y)),
\end{eqnarray}
\begin{equation}
\label{G55b}
-3b_0^2e^{2A_0(y)}\left(\ada\!\!\!\!(t)+\adda(t)\right)
-3A^\prime_0(y)(3f^\prime(t,y)+g^\prime(t,y))=b_0^2\kappa^2
P^5(y,t).
\end{equation}
The $y$ component of the  contracted Bianchi identities is still of the
form (\ref{bianchiy})
and the energy conservation equations for the brane as well as bulk
matter take on the form of (\ref{bianchit}).

The jump conditions are
\begin{equation}
\label{fjump}
[f^\prime(0)]_-=-{{\kappa^2b_0}\over{3}}\rho_\star(t),
~~~~~~~~~~~~~[f^\prime(1)]_-={{\kappa^2b}\over{3}}\rho(t).
\end{equation}
\begin{equation}
\label{gjump}
[g^\prime(0)]_-=\kappa^2b_0(P_\star(t)+\frac{2}{3}\rho_\star(t)),
~~~~~~~~~~~~~[g^\prime(1)]_-=-\kappa^2b (P(t)+\frac{2}{3}\rho(t)).
\end{equation}
We may now solve for $f(t,y)$ and $g(t,y)$ as follows. As before the leading order Einstein equations
lead to $A_0(y)=kb_0y$. Following our exact solution in the previous
section, we assume the matter has the form (\ref{form}). 
Then integrating the  next to leading order  $G_{00}$ equation leads to
\begin{equation}
f(t,y)=A(t)e^{4kby}-\frac{1}{4k^2}\left[ \ada -\frac{\kappa^2}{3}\hat\rho_B(t)\right] e^{2kby}.
\end{equation}Imposing the jump conditions at 0 and 1, yields the 
relations
\begin{equation}
8k^2 A(t)-  \ada =-\frac{k\kappa^2}{3} \rho_\star(t),
\end{equation}
\begin{equation}
-8k^2 A(t)e^{4kb}+e^{2kb}\ada =-\frac{k \kappa^2 }{3} \rho(t),
\end{equation}
which leads to the FRW-like equation for the expansion rate 
\begin{equation}
\label{Ares}
\ada\!\!\!\!(t)=\frac{8\pi
G_N}{3}\left[\frac{\hat{\rho}_B}{k}(t)(1-e^{-2kb})+\rho_\star(t)+\rho(t)
e^{-4kb}\right].
\end{equation}
Where we have used the relation $8\pi G_N=\frac{\kappa^2k}{1-e^{-2kb_0}}$,
which can be read off of the effective four dimensional action.
To leading order, $\frac{\dot{a}}{a}$ is the 
expansion rate on the visible brane.
Notice that this result  agrees with the contribution to the expansion rate
 that one would get by averaging the
bulk energy over the extra dimension. As with the Planck brane matter
there
is no warp factor suppression, which may at first be alarming.
To determine whether or not this result presents a problem in these
models we need to know how the bulk energy density is
normalized. Obviously,
if it scales like $M_{pl}$, the cosmology is not viable. The correct
normalization will be calculated in section four.

We may now solve for $g(t,y)$ by assuming an equation of state
$\rho_B=w_B P_B$ and imposing the $G_{05}$ equation in conjunction with the
time component of the Bianchi identity.
 \begin{equation}
\frac{\dot{a}}{a} g(t,y)=-\frac{e^{2kb_0y}}{4k^2}\left[ \frac{d}{dt}
\left(\ada-
\frac{\kappa^2}{3}\hat{\rho}_B(t)\right) +\frac{\dot{a}}{a}  \left(   \ada -\frac{\kappa^2}{3} \hat{\rho}_B(t)\right)\right] +e^{4kb_0y}(\dot{A}(t)+\frac{\dot{a}}{a} A(t))
\end{equation}
Imposing the jump conditions for $g(y)$,  reproduces the usual 
energy conservations for the branes 
\begin{eqnarray}
\dot{\rho}(t)+3 \frac{\dot{a}}{a}(t)\left(P(t)+\rho(t)\right)&=&0\nonumber
\\
\dot{\rho}_\star(t)+3 \frac{\dot{a}}{a}(t)\left(P_\star(t)+\rho_\star(t)\right)=0.
\end{eqnarray}
Thus, to the order we are working, the branes expand at the
same rate.

Similarly the $G_{ii}$ Einstein equation leads to the FRW like
relation for the acceleration
\begin{equation}
\frac{\ddot{a}}{a}(t)=-\frac{4 \pi G_N}{3} \left[ 3(P_\star(t) +P(t) e^{-4 k b_0}+
\frac{\hat{P}_B}{k}(t)(1-e^{-2kb}))+\rho_\star(t) +\rho(t) e^{-4kb}+
\frac{\hat{\rho}_B}{k}(t)(1-e^{-2kb_0}) \right].
\end{equation}

The $G_{55}$ equation
leads to the requirement
\begin{equation}
4A(t)+\frac{a}{\dot{a}} \dot{A}(t)=0.
\end{equation}
However, upon inspection we see that $A(t)$ is independent of the bulk energy density  and pressure and that
the constraint from the $G_{55}$ equation is just the usual unphysical relation,
 which results from the 
lack of a stabilizing potential, $\rho_\star(t)=-e^{-4kb_0}\rho(t)$.
This is just a sign that the bulk fluid has generated no radion potential.
We shall see that this is a consequence of our simple ansatz for the metric, which
greatly restricts the form of the bulk energy density, and that
this solution will only be valid for a massless bulk field at temperatures
low enough as not to excite higher KK modes.

\section{The Bulk Energy Density}
The solution found in the previous section necessitated that
the bulk energy density and pressure have the particular form (\ref{form}).
As we mentioned previously,  while the constraint
is satisfied for a bulk cosmological constant we are more interested in the
case of a thermal fluid, as this is the case of phenomenological relevance.
We thus would like to calculate the thermal expectation value
\begin{equation}
\langle T_{A B}(t,y)\rangle=Tr [T_{AB}(t,y) e^{-\beta_0 H}],
\end{equation}
in our background solution to test for consistency. 
We have assumed that  the expansion rate is
small enough that we have an approximate time-like killing vector, 
 and that local thermodynamic equilibrium
can be established.
Furthermore, we normalize the time-like killing vector
such that the Hamiltonian generates time translations on the visible
brane. That is to say that, $T$ will correspond to the proper temperature
on the brane. Requiring maximization of the proper entropy leads to
the relation $\sqrt{g_{00}} T={\rm const}$. 
Thus, the proper temperature measured by a local observer will grow  with the warp factor as one moves away from the
visible brane. This growth prevents the flow of heat between regions of
differing gravitational potential. We will further assume that the temperature
is low enough so that the modes of interest are not strongly interacting.

Let us first consider the field
of interest to be, for now,  a scalar, whose Hamiltonian
will be
\begin{equation}
H=\frac{1}{2}e^{-2A_0(y)}(\dot{\phi}^2+(\vec{\nabla}\phi)^2)+\frac{1}{2b_0^2}(\phi^\prime)^2e^{-4A_0(y)}+\frac{1}{2}m^2 \phi^2 e^{-4A_0(y)}.
\end{equation}
We then perform the usual Kaluza-Klein decomposition
\begin{equation}
\phi(x,y)=\sum_n\phi_n(x)\frac{\psi_n(y)}{\sqrt{b_0}},
\end{equation}
where $\psi_n(y)$ satisfy
\begin{equation}
-\frac{1}{b^2}\frac{d}{dy}\left(e^{-4A_0(y)} \frac{d\psi_n}{dy}(y)\right)+
m_B^2e^{-4A_0(y)}\psi_n=m_n^2e^{-2A_0(y)}\psi_n(y), 
\end{equation}
and the inner product is defined as
\begin{equation}
\int^1_{-1}dy \psi_n(y) \psi_m(y)e^{-2A_0(y)}=\delta_{nm}.
\end{equation}
The eigenmodes are linear combinations of Bessel functions, and were
computed in \cite{GWII}.
The relevant components of the stress energy tensor are
\begin{eqnarray}
T_{00}=\sum_n \frac{1}{2 b_0}\psi_n(y)^2 T_{00}^n+\frac{1}{2}\phi_n^2(x)
 e^{2 A_0(y)}\frac{d}{dy}({{\psi_n(y)
\psi_n^\prime(y)}\over{b_0^3}}e^{-4A_0(y)}) \cr \nonumber \cr
T_{ii}=\sum_n \frac{a^2(t)}{2 b_0}\psi_n(y)^2
T_{ii}^n+\frac{1}{2}\phi_n^2(x) e^{2 A_0(y)}\frac{d}{dy}({{\psi_n(y)
\psi_n^\prime(y)}\over{b_0^3}}e^{-4A_0(y)})
\cr \nonumber \cr
T_{55}=\sum_n \frac{1}{2}\phi_n^2(x) e^{2 A_0(y)}\frac{d}{dy}({{\psi_n(y)
\psi_n^\prime(y)}\over{b_0}}e^{-4A_0(y)})+\frac{1}{2}\phi_n^2(x)\frac{(\psi_n^\prime(y))^2}{b_0}.
\end{eqnarray}
$T_{00}^n$ and $T_{ii}^n$ are the four dimensional energy density and
pressure for the $nth$ mode.
In these expressions we have dropped mixing terms which will not
contribute to the thermal average and have used the free field equations
of motion to simplify $T_{55}$.
$T_{05}$, which would give rise to a net flow onto the branes, has
 vanishing thermal expectation value.
Consider now the contribution from one, say the lowest, Kaluza-Klein
mode $n=0$ in the limit of vanishing bulk mass (a fine tuned case when
the
bulk is not supersymmetric).
In this limit, the lowest mode is massless with a constant wave function.
We find
\begin{equation}
\label{massless}
\rho_B(y,t)=e^{2A_0(y)}\psi^2_0(y)\frac{\pi^2}{30 b_0} T^4,
\end{equation}
where again $T$ is the proper temperature, as measured by a visible
brane observer. Note that, as one would expect, the local energy
density is weighted by the wave function and now has a profile
which will only be consistent with our solution if it is trivial.
The normalized
wave function is given by $\psi^2_0(y)=kb_0(1-e^{-2kb_0})$, and we see from 
eq.(\ref{massless}) that, up to exponentially suppressed terms, the zero mode contributes to the energy
density exactly like a field trapped to the brane, with its normalization
set by the AdS curvature scale. 
Thus, the solutions discussed in the previous sections are
valid for a massless bulk field in the limit where the temperature
is too small to excite the higher Kaluza-Klein modes.
It is straightforward in this case to see that the bulk pressure in the
direction transverse to the branes vanishes and that bulk pressure
just reduces to the four dimensional pressure in the same way that
the energy density does.

Before going on to the more interesting massive case, we point out that
a massless bulk field is interesting in that it leads to
no deviations from standard cosmology, which seems rather surprising at first.
The lowest mode of a massless bulk field generates  no bulk
dynamics at all. Furthermore, its contribution to the expansion rate
is {\it exactly} the same as in the standard cosmology. Contrast
this to a massless field trapped to the brane which has non-linear
contributions to the expansion rate, and which generates a shift
in the lapse function \cite{ClineII}. 

\section{The Massive Bulk Fields}
 Thus, we would like to extend our
previous results to the case of a massive mode. Again, to make the
problem tractable, we will assume that we are only exciting the
lowest eigenmode. In principle it is possible to treat the sum of
several modes without conceptual change. 
We will find solutions to Einstein's equations by perturbing around
the massless solution via an expansion in the
small parameter $m_B/k$. We will further assume that the bulk energy
density is small compared to the brane tension, as we did in the 
previous section. This 
will allow us to drop time derivatives since they are higher
order in $\rho$. 
Writing $\psi_n(y)=C(1+m_B^2 \delta(y))$  leads
to \begin{equation}
\label{KK}
\delta^{\prime \prime}(y)-4kb_0\delta ^{\prime}(y)=b_0^2(1-\frac{m_0^2}{m_B^2} 
e^{2kb_0 y}),
\end{equation}
whose normalized solutions are given by 
\begin{equation}
\psi(y)=kb_0 \left(1+ \frac{m_B^2}{8 k^2}(3+2e^{2kb_0y}-e^{2kb_0(2y-1)}(1-e^{-2kb_0})-4kb_0(1+y)),
 \right)
\end{equation}
where $m_0$ is the massive of the lowest lying mode, $m_B$ is the
bulk mass and  exponentially suppressed terms have been dropped.
We impose the boundary conditions 
$\psi^\prime(0)=\psi^\prime(1)=0$, since we have not included any interactions
of this field on the brane.  This assumption is rather unnatural since
quantum effects will always induce operators localized to the brane
\cite{Georgi,GWIII}. However, it is not difficult to modify this analysis
to include these effects, which only complicate the algebra and do not
lead to any conceptual changes. 
Imposing the boundary conditions  we find that the mass of the mode is
given by $m_0^2=\frac{m_B^2}{2}(1-e^{-2kb_0})$. This case is again 
a
 fine tuned case
since the natural scale for the bulk mass is order $M_{pl}$, but is 
sufficient for our purposes.
The average energy density is again just
the canonical four dimensional energy density for a thermal fluid, as
a consequence of the topological constraint arising from (\ref{KK}).

Since the bulk profile is now no longer trivial,
 we will have to 
generalize our naive ansatz. For later convenience we  choose the
slightly different form
\begin{equation}
\label{newform}
ds^2=e^{-2\alpha(y,t)}dt^2-a_0^2(t)e^{-2 \beta(y,t)}d\vec{x}^2-b(t,y)^2dy^2.
\end{equation}
Where 
\begin{eqnarray}
\alpha(y,t)=A_0(y)+\delta \alpha(y,t) \nonumber \cr
\beta(y,t)=A_0(y)+\delta \beta(y,t) \nonumber \cr
b(y,t)=b_0+\delta b(y,t).
\end{eqnarray}
All the variations are of order $O(\frac{m^2}{k^2}\frac{\rho}{V+0})$, so that their
time derivatives will be suppressed.
We may neglect the brane matter since their effects are controlled by
a different expansion parameter, and the linearized results from the
previous section may be carried over directly. 

It has been pointed out in \cite{ClineII} that the 
 Einstein equations simplify quite a bit, if we choose to work
in terms of the variables
\begin{equation}
\eta=\delta \beta^\prime(y,t) -(A_0^\prime(y))^2\frac{\delta b(y,t)}{b_0} ~~~~~
\sigma(y,t)=\delta \alpha^\prime(y,t)-\delta \beta^\prime(y,t),
\end{equation}
which are invariant under gauge transformations which leave the branes
invariant. In terms of these variables, the $G_{00}$ and $G_{55}$  Einstein equations are 
\begin{equation}
4A_0^\prime(y)\eta(y,t)-\eta^\prime(y,t)=-\frac{e^{2A_0(y)}}{3}\kappa^2 \delta T_{00}(y,t),
\end{equation}
\begin{equation}
A_0^\prime(y)(4 \eta(y,t)+\sigma(y,t))=\kappa^2 \delta T_{55}(y,t),
\end{equation}
and the combination $G_{00}+b_0^2 e^{2 A_0(y)} G_{ii}$ is given by
\begin{equation} 
\sigma^\prime(y,t)-4A_0^\prime(y) \sigma (y,t)=
-\kappa^2 e^{2 A_0(y)}b_0^2(\delta T_{00}(y,t)+\frac{\delta T_{ii}(y,t)}{a_0(t)^2}).
\end{equation}
We note that the average of the first two equations vanishes as a consequence
of the topological constraint imposed by the fact that we are working on a
compact manifold. This is a manifestation of the point that to linear
order in the perturbations there are no corrections to the averaged Einstein
equations \cite{CsakiII}. We have the further simplification that both the
energy density and bulk pressure have the same $y$ dependence so that
the solutions to the first two equations are identical up to an overall
factor.
We find
\begin{eqnarray}
&&\eta(y,t)=\frac{-b_0 \kappa^2 m_B^2}{12k^2}\rho_4(t) \left( -\left( e^{kb_0\,\left( 1 + 3y \right) }kb_0y     \left( e^{kb_0\left( 1 + y \right) } +
 2\cosh (kb_0\left(  y -1\right) ) \right) 
     \left( \coth (b_0k)-1 \right)  \right) +  \nonumber  \right. \cr
&&
\left. 
 e^{b_0k + 4b_0ky}\left(\coth (b_0k)-1 \right) \sinh (b_0k\left( y-1 \right) )
   \sinh (b_0ky)
  +e^{3b_0ky}\left( 2 + e^{2b_0k} \right) b_0k{\rm{csch}}(b_0k)^2\sinh (b_0ky) \nonumber  \right),
\end{eqnarray}
and
\begin{equation}
\sigma(y,t)=-3(1+{{P_4(T)}\over{\rho_4(T)}})\eta(y,t).
\end{equation}
In deriving these expressions we have dropped the contribution from
the derivative turns in the expression for the stress energy, as they
only contribute at higher orders in the mass expansion.
We may now check to see if the $G_{55}$ equation is satisfied. This is the
equation which accounts for the induced radion potential and leads
to, in the case of brane matter,  the unphysical constraint $\rho=-e^{-4kb_0}\rho_\star$.
To the order we are working $T_{55}$ vanishes and the constraint equation
leads to the condition
\begin{equation}
\rho_4(T)=3P_4(T),
\end{equation} 
which is of course not possible for a massive thermal fluid.
Thus, we see that, at least with no external stabilization mechanism, 
there are no physical solutions with a thermal bulk fluid. It is simple to
include the effects of matter on the branes since their effect is controlled
by a separate expansion parameter. Since the Einstein equations are local
and the constraint on the brane matter is independent of the fifth dimension, 
there is no way to even fine tune the relation between the brane and bulk
matter to get a solution.

\section{Inclusion of a GW mechanism}

To determine the effects of a bulk thermal fluid, we now include
an additional scalar $\xi$, which is responsible for stabilizing the space.
This stabilization is accomplished by including a bulk as well
as brane potentials for $\xi$. An exact solution to this system without
matter was found in \cite{DeWolfe}. Here we will perturb around this
solution, and following \cite{ClineII}, choose the simple bulk potential 
$V(\xi)=\frac{1}{2}m_\xi^2 \xi^2$ and brane potentials 
$V_i(\xi)=m_i(\xi-v_i)^2$. The solution to the unperturbed coupled
Einstein scalar field equations are approximately given by
\begin{equation}
\xi_0(y)=v_0e^{-\epsilon k b_0 y} ~~~~~~~~A_0(y)=k b_0y-
\frac{\kappa^2}{12}v_0^2(1-e^{-2\epsilon b_0 y}),
\end{equation}
where $\epsilon\approx \frac{m_\xi^2}{4k^2}$.
A large hierarchy, without fine-tuning, is obtained by choosing
$e^{-kb_0}=(v_1/v_0)^{\epsilon^{-1}}$. 
The $G_{00}$ and $G_{ii}$ Einstein equations are unchanged, except
for the fact that gauge invariant variable is now
\begin{equation}
\eta(y,t)=\delta \alpha^\prime(y,t) -A_0^\prime(y) \frac{\delta b(y,t)}{b_0}-
\frac{\kappa^2}{3}\xi_0^\prime(y) \delta \xi(y,t),
\end{equation}
where $\delta \xi$ is the deviation induced by the mass of the
bulk matter. Note that there are still no jumps on the branes.
That is, in these Einstein equations there are no implied delta functions
induced by the shift in the fields, as they have all canceled
once the leading order equations have been used.

The $G_{55}$ equation is now given by
\begin{equation} 
A_0^\prime(y)(4\eta(y,t)+\sigma(y,t))+\frac{\kappa^2}{3}
\left(\xi_0^{\prime \prime}(y)\delta \xi(y,t)-\xi_0^\prime(y) 
\delta \xi^\prime(y,t)+(\xi_0^\prime (y))^2\frac{\delta b(y,t)}{b_0}\right)=0+ O(m_B^4/k^4).
\end{equation}
Given that we have already solved the other Einstein equations we can solve
the entire system by making the simplifying assumption that the brane potentials
are stiff, so that the fluctuations of $\xi$ on the branes can be taken to
be zero. We may then use residual gauge invariance to eliminate 
$\delta \xi$ in the bulk as well.
Doing so allows us to calculate the shift in the radion due to the
bulk field. We find
\begin{equation}
2\int^1_0 dy \frac{\delta b(y,t)}{b_0}=\frac{m_B^2}{24 k^2}{{e^{2kb_0(2+\epsilon)}}\over{k^2v_0^2 b_0\epsilon^2}}(\rho_4(T)-3P_4(T))\approx \frac{m_B^2}{18k^2 m_r^2 }
{{(\rho_4(T)-3P_4(T))}\over{kb_0 M_{Pl}^2 e^{-2kb_0}}}.
\end{equation}
$m_r$ is the mass of the radion which in the limit we are working
is approximately given by $m_r^2=\frac{4}{3}\kappa^2(\epsilon v_0k)^2e^{-2(kb_0+\epsilon)}$ \cite{CsakiII,radion}. Note that the bulk fluid again behaves like a brane fluid, 
which couples conformally to the radion.

The more interesting quantity is the change in the lapse function as this
quantity tells us how the weak scale changes with time.
Writing
\begin{equation}
M_W(t)/M_W(t_0)\approx e^{-2 \int_0^1 \delta \alpha^\prime(y,t)dy}
\end{equation}
We find
\begin{equation}
M_W(t)/M_W(t_0)\approx 1-\frac{m_B^2}{18k^2}{{\rho_4(T)}\over{M_{pl}^2k^2 e^{-4kb_0}}}(1+\frac{3P_4(T)}{2\rho_4(T)})+\frac{\delta b(t)}{b_0}.
\end{equation}
Notice that in the limit where $m_0<T<m_1$ (which can be attained by fine tuning
the bulk mass), and where our results are still valid, 
the contribution from the radion shift is subleading. 

\section{Conclusions and Outlook}

In this paper we have calculated perturbations away from the 
vacuum $AdS_5$ compactified
geometry due to the presence of a thermal bulk in the limit where
$T<m_{KK}$.  We have 
shown that a bulk fluid contributes to the expansion rate 
in a similar fashion to a fluid which is confined to the visible brane and
that the induced dynamics of the fifth dimension are controlled by the mass 
of the bulk field. The existence of the bulk fluid does nothing to ameliorate
the usual fine tuning between the brane matter when there is no stabilizing
potential. In fact, without a stabilizing potential there are no physical
solutions, at least that can be found by perturbing 
around the vacuum solution, 
when a thermal bulk fluid is introduced. 
Upon introducing a stabilizing potential we found that
the shift in the weak
scale due to the existence of the bulk fluid is parametrically 
suppressed relative to the
contribution from a brane by $m_B^2/k^2$. This stems from the fact that all
the bulk dynamics induced by the bulk fluid are controlled by this
parameter. Indeed, 
 we may conclude that the onset
of dynamics in the fifth dimension, due to a bulk field,
 is triggered at a temperature
near the mass of the first mode which has a non-trivial vacuum
profile.

Our results are only valid in the limit where $T<m_{KK}^{(2)}$,
where $m_{KK}^{(2)}$ is the mass of the second Kaluza Klein
excitation, so that contributions from higher excitations are
suppressed.  
It would be interesting to find solutions where the
mass of the first mode approaches the TeV scale, so that there
is no need for fine tuning. This would
entail solving the Einstein equations with the full Bessel function
wave function source. Furthermore, one could then also include more
Kaluza-Klein excitations. This would allow for a study
at temperatures high enough that the free energy begins to
scale like $T^5$.
 Presumably, as the temperature increases further, the local
energy density near the Planck brane 
will approach the Planck scale (this can't be 
seen from our results since our calculation of the 
energy density was only valid for
smaller temperatures where we could do perturbation theory).
Once such energy densities are reached, black-hole 
formation seems inevitable
and the change in geometry to AdS-Schwarzschild
marks a phase transitions in the dual conformal field theory.

\acknowledgments 
The author is  indebted to Walter Goldberger, Ben Grinstein,
   Rich Holman and Ted Jacobson    
for helpful discussions.
This work was supported in part by the Department of Energy under
grant numbers DOE-ER-40682-143.

\tighten


\end{document}